\documentclass{article}

\usepackage{PRIMEarxiv}

\usepackage[utf8]{inputenc} 
\usepackage[T1]{fontenc}    
\usepackage{hyperref}       
\usepackage{url}            
\usepackage{booktabs}       
\usepackage{amsfonts}       
\usepackage{nicefrac}       
\usepackage{microtype}      
\usepackage{lipsum}
\usepackage{fancyhdr}       
\usepackage{graphicx}       
\graphicspath{{media/}}     

\usepackage{booktabs}

\usepackage{multirow}
\usepackage[skins]{tcolorbox}

\usepackage{subcaption}
\usepackage{xspace}
\usepackage{graphicx}
\usepackage{array}
\usepackage{xcolor}
\usepackage{bm}
\usepackage{amsmath}
\usepackage{hyperref}

\usepackage{tikz}
\usepackage{xcolor}

\usepackage{float}
\newtcolorbox{custombox}[1]{
	colback=gray!10,
	colframe=gray!20,
	left=1mm,
	right=1mm,
	top=1mm,
	bottom=1mm,
	fonttitle=\bfseries,
	arc=2mm,
	leftrule=0mm,
	rightrule=.5mm,
	toprule=0mm,
	bottomrule=.5mm,
	notitle,
	before=\par\smallskip\noindent,
	before upper={\textbf{#1: } },
}

\newsavebox\CBox

\newcommand{\deepscenario}{$\mathrm{DeepScenario}$\xspace}

\newcommand{\gpt}{\textit{GPT-3.5}\xspace}

\newcommand{\llama}{\textit{Llama2-13B}\xspace}

\newcommand{\mistral}{\textit{Mistral-7B}\xspace}

\newcommand{\robustnessScore}{\textit{RS}\xspace}
\newcommand{\topcount}{\textit{T-5C}\xspace}
\newcommand{\bottomcount}{\textit{B-5C}\xspace}
\newcommand{\inconsistentcount}{\textit{IC}\xspace}

\newcommand{\finding}[1]{
\begin{center}
    \fcolorbox{black}{gray!10}{\parbox{.96\columnwidth}{#1}}
\end{center}
}

\usepackage{adjustbox}
\usepackage{listings}
\usepackage{color}
 
\definecolor{dkgreen}{rgb}{0,0.6,0}
\definecolor{gray}{rgb}{0.5,0.5,0.5}
\definecolor{mauve}{rgb}{0.58,0,0.82}
\definecolor{gray}{rgb}{0.4,0.4,0.4}
\definecolor{darkblue}{rgb}{0.0,0.0,0.6}
\definecolor{lightblue}{rgb}{0.0,0.0,0.9}
\definecolor{cyan}{rgb}{0.0,0.6,0.6}
\definecolor{darkred}{rgb}{0.6,0.0,0.0}

\lstdefinelanguage{XML}
{
  morestring=[s][\color{mauve}]{"}{"},
  morestring=[s][\color{black}]{>}{<},
  morecomment=[s]{<?}{?>},
  morecomment=[s][\color{dkgreen}]{<!--}{-->},
  stringstyle=\color{black},
  identifierstyle=\color{lightblue},
  keywordstyle=\color{red},
  morekeywords={xmlns,xsi,noNamespaceSchemaLocation,type,id,x,y,source,target,version,tool,transRef,roleRef,objective,eventually}
}

\title{Reality Bites: Assessing the Realism of Driving Scenarios with Large Language Models}

\author{
  Jiahui Wu \\
  Simula Research Laboratory and \\ University of Oslo \\
  Oslo, Norway \\
  \texttt{jiahui@simula.no} \\
   \And
  Chengjie Lu \\
  Simula Research Laboratory and \\ University of Oslo \\
  Oslo, Norway \\
  \texttt{chengjielu@simula.no} \\
   \And
  Aitor Arrieta \\
  Mondragon University \\
  Mondragon, Spain \\
  \texttt{aarrieta@mondragon.edu} \\
   \And
  Tao Yue \\
  Beihang University \\
  Beijing, China \\
  \texttt{yuetao@buaa.edu.cn} \\
   \And
  Shaukat Ali \\
  Simula Research Laboratory \\
  Oslo, Norway \\
  \texttt{shaukat@simula.no} \\
}

\begin{document}
\maketitle

\begin{abstract}
Large Language Models (LLMs) are demonstrating outstanding potential for tasks such as text generation, summarization, and classification. Given that such models are trained on a humongous amount of online knowledge, we hypothesize that LLMs can assess whether driving scenarios generated by autonomous driving testing techniques are realistic, i.e., being aligned with real-world driving conditions. To test this hypothesis, we conducted an empirical evaluation to assess whether LLMs are effective and robust in performing the task. This reality check is an important step towards devising LLM-based autonomous driving testing techniques. For our empirical evaluation, we selected 64 realistic scenarios from \deepscenario--an open driving scenario dataset. Next, by introducing minor changes to them, we created 512 additional realistic scenarios, to form an overall dataset of 576 scenarios. With this dataset, we evaluated three LLMs (\gpt, \llama, and \mistral) to assess their robustness in assessing the realism of driving scenarios. Our results show that: (1) Overall, \gpt achieved the highest robustness compared to \llama and \mistral, consistently throughout almost all scenarios, roads, and weather conditions; (2) \mistral performed the worst consistently; (3) \llama achieved good results under certain conditions; and (4) roads and weather conditions do influence the robustness of the LLMs.
\end{abstract}

\keywords{Large Language Models \and Realistic Driving Scenarios \and Robustness}

\section{Introduction}

The autonomy of vehicles has significantly increased in the last few years, up to a level where human intervention is not even required for some vehicles in certain restricted situations. To achieve such a level, it is paramount to ensure the high dependability of autonomous driving systems (ADSs) through automated testing. Because of this, significant effort has been devoted by the research community to devise effective and efficient techniques to test ADSs~\cite{nalic2020scenario,10.1145/3579642,10.1109/TSE.2022.3170122}.
To this end, automated test case generation aims to generate test cases (e.g., driving scenarios in our case) to fail ADSs, for instance, by finding collisions. To this end, different techniques have been proposed in the last few years, including search-based techniques~\cite{abdessalem2018testing,abdessalem2018ASE,calo2020generating,li2020av}, reinforcement learning techniques~\cite{feng2023dense,9468363,lu2022learning,10.1109/ICSE48619.2023.00155} and causality-driven techniques~\cite{giamattei2023causality}.

Unfortunately, such techniques produce unrealistic driving scenarios, especially when simulated environments are involved~\cite{10.1109/TSE.2022.3202311}.
For example, the simulation may not precisely represent the dynamic characteristics of vehicles during a collision realistically, such as how they respond to impact forces, or how they affect each other's trajectories.

Different approaches have been proposed to overcome such problems, such as finding ADS configurations that avoid such situations and gain confidence in the validity and realism of the generated driving scenarios for testing~\cite{calo2020generating}. Besides, some ADS testing techniques particularly aim to generate realistic driving scenarios with measures such as enforcing realistic constraints~\cite{lu2023deepqtest} and utilizing real-world driving data~\cite{zhang_risk_2022,yan2023learning,gambi2019generating}.
Those techniques require compute-intensive approaches combining search algorithms with simulation-based testing that usually involves a simulated environment, while, as Stocco et al.~\cite{10.1109/TSE.2022.3202311} indicate, there exists a reality gap between the simulation-based test and real-world test.
In practice, it is extremely time-consuming to assess whether the generated test scenario is realistic for the real-world testing setup due to the complexity and infinite number of driving scenarios. 
Subsequently, developing effective and efficient techniques to assess whether a driving scenario for testing an ADS is realistic is essential. 


Large Language Models (LLMs) have demonstrated great potential in various domains including context understanding, machine translation, and logic reasoning. With the capabilities of open-world cognition and common-sense reasoning, LLMs have gained significant attention in the domain of ADSs such as decision-making and environmental awareness~\cite{yang2023llm4drive,cui2023survey}.
Given that LLMs have learned a large amount of information, we conjecture that they can provide adequate means to assess the realism of ADS driving scenarios. To validate our hypothesis, we carry out a comprehensive empirical evaluation to assess the robustness of three prominent LLMs, i.e., \gpt~\cite{openaiGPT3.5}, \llama~\cite{touvron2023llama}, and \mistral~\cite{jiang2023mistral} in assessing the realism of driving scenarios. We validated the selected LLMs on 64 realistic scenarios (original) from open-source driving scenario set \deepscenario~\cite{10174023}. Moreover, we generated 512 additional scenarios by mutating the original scenarios. As a result, we have a dataset of 576 realistic scenarios. We defined \textit{robustness score} to measure LLM's robustness and evaluated LLMs by considering four independent variables, namely road, weather, scenario, and parameter (i.e., the parameter settings of the vehicle, such as position). 
Our results show that: (1) Overall, \gpt achieves the highest \textit{robustness score} (i.e., 12.59 out of 20) compared to \llama and \mistral, whose \textit{robustness scores} are 9.48 and 5.60, respectively. (2) Regarding the robustness by roads, \gpt has the most robust performance on 3 out of 4 roads (i.e., \textit{robustness scores} with 12.64, 12.20, and 12.38), whereas \mistral is the worst for all roads. (3) Concerning the robustness to weather conditions, a similar pattern to that for roads can be observed, i.e., \gpt achieves the best \textit{robustness scores} on 3 out of 4 weather conditions (i.e., 12.65, 11.34, and 13.56), whereas \mistral is the worst for all weather conditions.



In summary, we conducted an empirical study to evaluate LLMs' abilities in assessing the realism of driving scenarios with three LLMs and an open-source driving scenario dataset. We presented a novel metric, i.e., \textit{robustness score}, to measure the robustness of LLMs for realistic scenario identification. We analyzed the impact of different independent variables (i.e., road, weather, parameter, and scenario) on the robustness of the LLMs and reported our findings. Such findings are a stepping stone for developing novel testing strategies to generate realistic driving scenarios with LLMs. 


\section{Related Work}\label{sec:related_work}

Scenario-based ADS testing cost-effectively generates realistic (if possible) driving scenarios to test how well an ADS interacts with the environment under these scenarios. Various ADS testing approaches have been proposed, e.g., based on search-based testing (SBT)~\cite{abdessalem2018testing,abdessalem2018ASE,calo2020generating,9251068,10.1145/3551349.3556897}, reinforcement learning (RL) based testing~\cite{feng2023dense,9468363,lu2022learning,lu2023deepqtest,10.1109/ICSE48619.2023.00155} and causality-driven testing~\cite{giamattei2023causality}. SBT-based approaches use fitness functions to guide optimized scenario generation. For instance, FITEST~\cite{abdessalem2018ASE} uses a multi-objective search algorithm to generate critical driving scenarios, whereas AV-FUZZER~\cite{9251068} combines fuzzing and SBT to identify safety violations. 
RL-based approaches focus on adaptive scenario generation in dynamic driving environments. 
For example, DeepCollision~\cite{lu2022learning} generates critical scenarios by dynamically configuring an ADS's operating environment, whereas MORLOT~\cite{10.1109/ICSE48619.2023.00155} combines RL and multi-objective search, for violating multiple requirements simultaneously. 
Some ADS testing techniques particularly aim to generate realistic driving scenarios with measures, e.g., enforcing realistic constraints~\cite{lu2023deepqtest} and using real driving data~\cite{zhang_risk_2022,yan2023learning,gambi2019generating,feng2023dense}. However, these techniques involve a virtual environment in the testing loop, and evidence showed a reality gap between virtual and physical ADS testing~\cite{10.1109/TSE.2022.3202311}. Thus, it is important to assess whether driving scenarios generated with these techniques are realistic to be used in real-world ADS testing. Given that LLMs are trained on an extensive amount of real-world knowledge, we aim to evaluate LLMs' abilities in assessing the realism of scenarios generated by ADS testing techniques. This reality check is a critical step towards developing LLM-based realistic driving scenario generation for ADS testing.


Advances in LLMs provide possibilities to use them for solving diverse software engineering (SE) problems~\cite{zhang2023survey,fan2023large,hou2023large}, e.g., code generation, completion, and repair~\cite{mastropaolo2021studying,wei2023magicoder,10172854,10179324}, requirement completion and classification~\cite{luo2022prcbert,luitel2023improving}, and software testing~\cite{wang2023software,10172800,liu2023testing}. For code generation and completion, various LLMs, e.g., \gpt have been applied to interpret natural language descriptions and generate code snippets. Mastropaolo et al.~\cite{mastropaolo2021studying} empirically investigated how the pre-trained and fine-tuned T5 model~\cite{raffel2020exploring} perform to support code-related tasks such as automated bug-fixing and code summarization. OSS-INSTRUCT~\cite{wei2023magicoder} prompts LLMs with open-source code snippets to generate high-quality coding instruction data to boost LLMs' performance via instruction tuning.
Evidence also shows that LLMs exhibit strong contextual understanding, enabling them to capture nuances and intricacies of natural language requirement~\cite{zhang2023preliminary,fan2023large}. 
PRCBERT~\cite{luo2022prcbert} is a prompt learning approach for requirement classification using BERT-based pre-trained LLMs.
Furthermore, LLMs are being explored to generate test cases based on natural language descriptions or existing code, thereby automating software testing ~\cite{wang2023software}. CodeMosa~\cite{10172800} integrates SBT and LLMs for code (i.e., CodeX~\cite{chen2021evaluating}) to generate high-coverage test cases for programs under test. Additionally, studies have investigated LLMs characteristics, e.g., as non-determinism~\cite{ouyang2023llm} and uncertainty~\cite{huang2023look}. Notably, empirical evaluations predominantly concentrate on SE tasks related to code generation, question answering, and text summarization. Our work is the very first to explore LLMs' potential in assessing the realism of driving scenarios of ADSs.


LLMs in the ADS context primarily focus on ADS control design policies and environment understanding~\cite{wu2023language,ding2023hilmd,keysan2023text} and decision-making~\cite{liu2023mtdgpt,sha2023languagempc,wen2023dilu,fu2024drive}. 
LLM's few-shot learning enables faster and more accurate learning and reasoning~\cite{p2023protoclip,lin2023multimodality}. In the perception phase of autonomous driving, such capabilities can greatly improve ADS's performance in understanding the dynamic environment. PromptTrack~\cite{wu2023language} predicts and tracks multiple 3D objects in driving environments with a given language prompt. HiLM-D~\cite{ding2023hilmd} incorporates high-resolution information (e.g., images sampled from videos) into multi-modal LLMs for risk object localization and intention and suggestion prediction tasks. 
LLM's open-world cognitive and commonsense reasoning can enhance ADS's decision-making capability~\cite{yang2023llm4drive,cui2023survey}. Liu et al.~\cite{liu2023mtdgpt} formulated the multi-task decision-making problem as a sequence modeling problem and then proposed MTD-GPT to manage multiple driving tasks simultaneously. DriveGPT4~\cite{xu2023drivegpt4} is an end-to-end ADS based on multi-modal LLMs that can interpret, reason, and answer diverse questions from human drivers to enhance ADS's driving ability. 
Regarding applying LLMs for ADS testing, TARGET~\cite{deng2023target} uses an LLM to parse traffic rules and generate executable scenario scripts. TARGET was evaluated on four ADSs, resulting in 217 driving scenarios, in which approximately 700 traffic rule violations, collisions, and other significant issues, including navigation failures, were identified. 
To compare with our work, TARGET aims to extract knowledge from traffic rules and generate scenario representations, but not to check scenario realism. In conclusion, some works use LLMs to develop control policies and generate driving scenarios, but there is no work studying the LLMs robustness in assessing the realism of driving scenarios, which is an important step towards building ADS testing techniques for generating realistic driving scenarios. To this end, it is necessary to conduct a comprehensive evaluation to see whether LLMs are effective and robust in assessing the realism of driving scenarios, which is our focus.


\section{Experiment Design}\label{sec:design}

We present the design of our empirical evaluation in this section. In detail, we describe the benchmark dataset used for our experiment in Section~\ref{subsec:benckmark_dataset}. Section~\ref{subsec:experiment_settings} presents the detailed experimental settings including the subject LLMs, temperature settings for LLMs, and the prompt design. We show our research questions in Section~\ref{subsec:RQs} and define the evaluation metrics in Section~\ref{subsec:metrics}.

\subsection{Benchmark Dataset}\label{subsec:benckmark_dataset}
To study LLMs' robustness in assessing the realism of driving scenarios, we created a benchmark dataset containing realistic scenarios from \deepscenario~\cite{10174023}. \deepscenario is an open dataset containing realistic driving scenarios generated with DeepQTest~\cite{lu2023deepqtest}, a reinforcement learning-based strategy. DeepQTest ensures that all scenarios in \deepscenario are realistic by benefiting from the realistic constraints in its scenario generation process and by introducing real-world weather data. For instance, when introducing a new object (e.g., a pedestrian or a vehicle) into the simulated environment, the object should keep a safe distance from other objects, restricting where the object can be generated with its initial position. 
Besides, DeepQTest enables the introduction of real-world weather conditions (from OpenWeather~\cite{openweather}, an open-source online weather database) into simulations to generate realistic scenarios. 


A driving scenario $S$ describes the temporal development of scenes, each of which describes a snapshot of the environment~\cite{ulbrich2015defining}. In \deepscenario, $S$ is defined as a tuple with scenes: $S$ = <$scene_1$, $scene_2$, ..., $scene_n$, T>, where $T$ and $n$ are the period that $S$ spans and the total number of scenes in $S$, respectively. A scene is defined as a 3-tuple: $scene$ = $<$\textit{ego story}, \textit{obstacle story}, \textit{environment state}$>$, which, respectively, describe operations and kinetic parameters of the ADS, static and dynamic obstacles, weather conditions, and the time of the day.
Listing~\ref{lst:scenario} shows a snippet example of a driving scenario in \deepscenario, which shows that the ego and non-player character (NPC) vehicles are characterized by four parameters, i.e., position, rotation, velocity, and angular velocity. The information in <$wayPoint$ $timeStamp=``i"$> indicates the state of the ego vehicle or NPC vehicle in the \textit{i-th} scene, where the time interval between each scene is 0.5 seconds.


We create the benchmark dataset by selecting realistic scenarios from \deepscenario by considering four different roads (i.e., R1 to R4) with various road structures, e.g., their lane structures, traffic signs and signals, crosswalks, and intersections, four weather conditions (rainy day (RD), rainy night (RN), sunny day (SD), clear night (CN)), and four driving scenarios containing an ego vehicle and one NPC vehicle. We select scenarios with only one NPC vehicle since more NPC vehicles will increase the prompt length to more than the maximum number of tokens (i.e., 4K) for \llama and \mistral, hence making outputs incomplete or not meaningful.
As a result, we collected 64 realistic scenarios (4 roads $\times$ 4 weather conditions $\times$ 4 driving scenarios) from \deepscenario. 

To create realistic variations to the realistic scenarios, we employ a mutation operator~\cite{laurent2023parameter}, which mutates values of the four parameters of the NPC vehicle in a scenario, i.e., position, rotation, velocity, and angular velocity, by 1\% increment (e.g., mutated position = original position * 1.01) or decrement (e.g., mutated position = original position * 0.99), in each timestamp of a selected realistic scenario. We mutate one parameter at a time in each timestamp, e.g., only increasing the value of the position of an NPC vehicle by 1\%, to produce one variation of the realistic scenario. As a result, we obtain 64 $\times$ 4 $\times$ 2 $=$ 512 variations to the 64 realistic scenarios.
We then manually checked all 512 scenarios by replaying them in the simulation and ensured their realism was still preserved. 
\lstset{language=XML}
\begin{lstlisting}[caption={Examplifying driving scenarios in DeepScenario},label=lst:scenario,basicstyle=\scriptsize]
<DeepScenario timestep="0.5">
		<Story name="Default">
			<ObjectAction name="Act_Ego0">
				<objectRef objectRef="Ego0"/>
				<WayPoint timeStamp="1">
					<DynamicParameters>
                            ......
						<Velocity velocityX="-0.0" velocityY="-0.25" velocityZ="-0.0"/>
						<AngularVelocity angularVelocityX="-0.001" angularVelocityY="-0.0" angularVelocityZ="0.001"/>
					</DynamicParameters>
					<GeographicParameters>
						<ObjectPosition positionX="-201.879" positionY="10.27" positionZ="217.72"/>
						<ObjectRotation rotationX="0.01" rotationY="180" rotationZ="-0.004"/>
					</GeographicParameters>
				</WayPoint>
    			<WayPoint timeStamp="2">
                    ......
				</WayPoint>
                    ......
    		</ObjectAction>
			<ObjectAction name="Act_NPC0">
				<objectRef objectRef="NPC0"/>
				<WayPoint timeStamp="1">
                    ......
				</WayPoint>
                    ......
    			</ObjectAction>
		</Story>
	</StoryBoard>
</DeepScenario>
\end{lstlisting}
\vspace{-13pt}

\subsection{Experimental Settings}\label{subsec:experiment_settings}
\label{sec:exper_settings}
We discuss various experimental settings for our empirical evaluation in this subsection. 

\begin{figure*}[!t] 
    \centering
    \includegraphics[width=\linewidth]{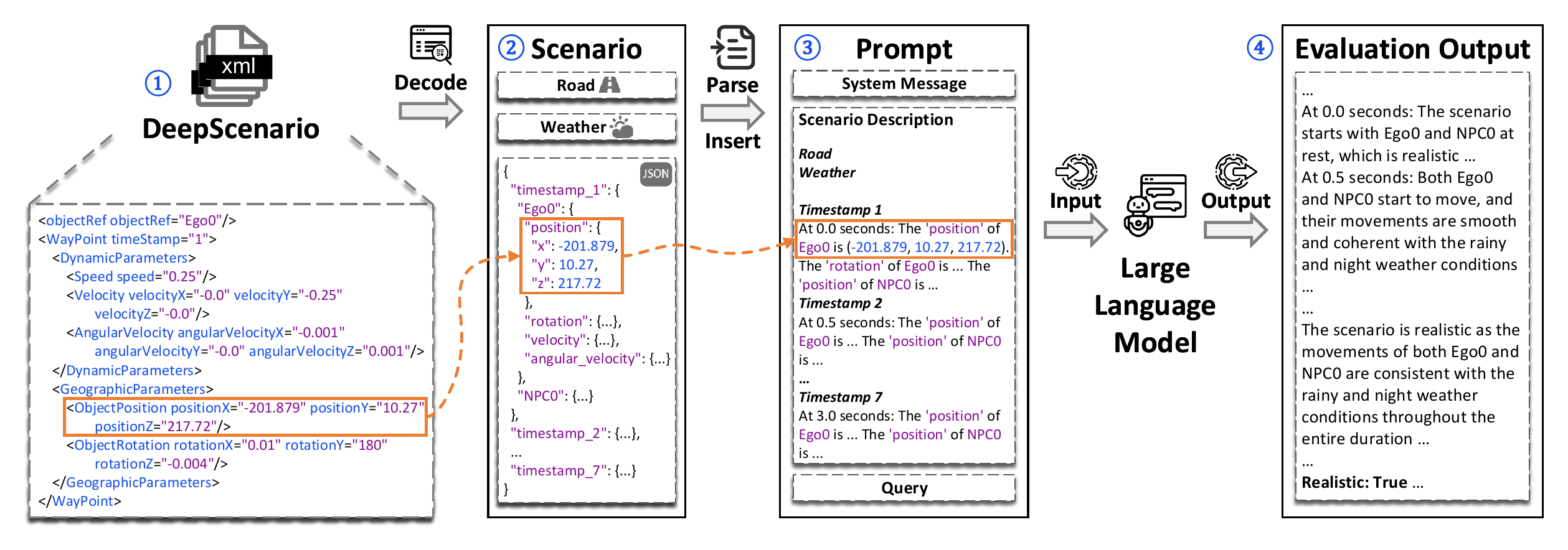}
    \caption{Process for generating a prompt from a DeepScenario scenario to evaluating the realism of the scenario with an LLM}
    \label{fig:overview}
\end{figure*}

\vspace{3pt}
\noindent\textbf{Subject LLMs.} We selected three representative LLMs: \gpt~\cite{openaiGPT3.5}, \llama~\cite{touvron2023llama}, and \mistral~\cite{jiang2023mistral}, for our evaluation by considering their availability, diversity, and performance under various general-purpose tasks. \gpt is a generative pre-trained transformer (GPT) developed by OpenAI. \llama is the pre-trained and fine-tuned LLM with 13 billion parameters created by Meta. 
\mistral is a 7-billion-parameter LLM built by Mistral AI. We selected \mistral since it showed better performance than \llama on various benchmarks including reasoning, mathematics, and code generation~\cite{jiang2023mistral}.

\vspace{3pt}
\noindent\textbf{Temperature and Repetitions.} Different temperature settings influence the quality of the task performed by an LLM~\cite{xu2022systematic}. Generally, a higher value enables an LLM to produce more diverse results but increases the non-determinism, whereas a lower value produces more deterministic results. In our context, we are interested in identifying realism more deterministically; therefore, we set a temperature setting of 0.0 for the three LLMs. Finally, LLMs are stochastic (even when the temperature setting is 0~\cite{ouyang2023llm}); therefore, to account for randomness, we asked each LLM to identify the realism of each scenario ten times.

\vspace{3pt}
\noindent\textbf{Prompt Design.} Figure~\ref{fig:overview} shows how various parameters are parsed from one driving scenario in our benchmark dataset to generate a prompt. It also shows the output generated by an LLM in response to such a prompt. In \textcircled{1}, at each timestamp, we extract specific variables and their values for each object in one driving scenario from \deepscenario, i.e., position, rotation, velocity, and angular velocity of the ego vehicle and NPC vehicle. We also obtain the corresponding roads and weather conditions for the driving scenario. Combining these variables, we organize the scenario information with a fixed format as shown in \textcircled{2}. Next, in \textcircled{3}, we use the scenario information to instantiate our fixed prompt template to generate the specific prompt used for LLM evaluation. Finally, we get the evaluation output from the LLM, as shown by \textcircled{4}.

For our empirical evaluation, we design one fixed prompt template including the system message, the scenario description, and the query for LLMs (see Table~\ref{tab:promptTemplate}). In the scenario description paragraph, we also provide examples.
\begin{table*}[!t]
\centering
\caption{Prompt template with example}
\label{tab:promptTemplate}
\resizebox{\linewidth}{!}{
\begin{tabular}{m{1.9cm}m{1.5cm}m{8cm}m{8cm}}
\toprule
\multicolumn{2}{l}{\textbf{Paragraph}} & \textbf{Template} & \textbf{Example} \\ \midrule
\multicolumn{2}{l}{System Message} & \multicolumn{2}{l}{\begin{tabular}{p{16cm}}… You are a helpful scenario realism evaluation assistant. You will evaluate the following autonomous driving scenario, and check whether the scenario is realistic …\end{tabular}} \\ \midrule
\multirow{9}{*}{\begin{tabular}{l}Scenario\\Description\end{tabular}} & Road & In the following scenario, Ego0's driving intention is to [\textit{Ego’s driving intention on the current road}]. & In the following scenario, Ego0's driving intention is to first drive on a dual-way road, then turn right to a one-way road of four lanes. \\ \cmidrule(r){2-4} 
 & Weather & The weather in this scenario is [\textit{weather conditions in the current scenario}]. & The weather in this scenario is sunny and day. \\ \cmidrule(r){2-4} 
 & Parameter & The [\textit{the parameter name}] of [\textit{the specific object in the current scenario}] is [\textit{the parameter triple values}]. & The `position' of Ego0 is (-293.259, 10.204, 1.007). \\ \cmidrule(r){2-4} 
 & Timestamp & At [\textit{the simulation time corresponding to the current timestamp}] seconds: [\textit{parameter descriptions of the Ego and NPC corresponding to the current timestamp}]. & At 0.0 seconds:
The `position' of Ego0 is (-293.259, 10.204, 1.007). … The `position' of NPC0 is (-273.203, 10.208, -6.955). … \\ \midrule
\multicolumn{2}{l}{Query} & \multicolumn{2}{l}{\begin{tabular}{p{16cm}}Your task is to perform the following actions … Evaluate the realism of the scenario … Output whether the scenario is realistic … Use the following format … Evaluation of the Realism of the scenario: <evaluation results> … Realistic: <True or False> …\end{tabular}} \\ \bottomrule
\end{tabular}
}
\end{table*}
Note that we provide a prompt with all four parameters (i.e., position, rotation, velocity, and angular velocity) of all objects (i.e., one ego and one NPC) in a scenario. Moreover, each LLM has a different default prompt format (e.g., \llama has the special \textit{<<SYS>>} token); thus, for each LLM we keep the main contents of the prompt same to ensure a fair comparison with other LLMs, but adapt it for its specific default prompt format. 
We generated 576 prompts to assess 64 realistic scenarios and 512 mutated scenarios. Each prompt is repeated 10 times, resulting in 5760 prompt requests per LLM.

\begin{figure}[!htbp]
    \centering
    \includegraphics[width=0.65\linewidth]{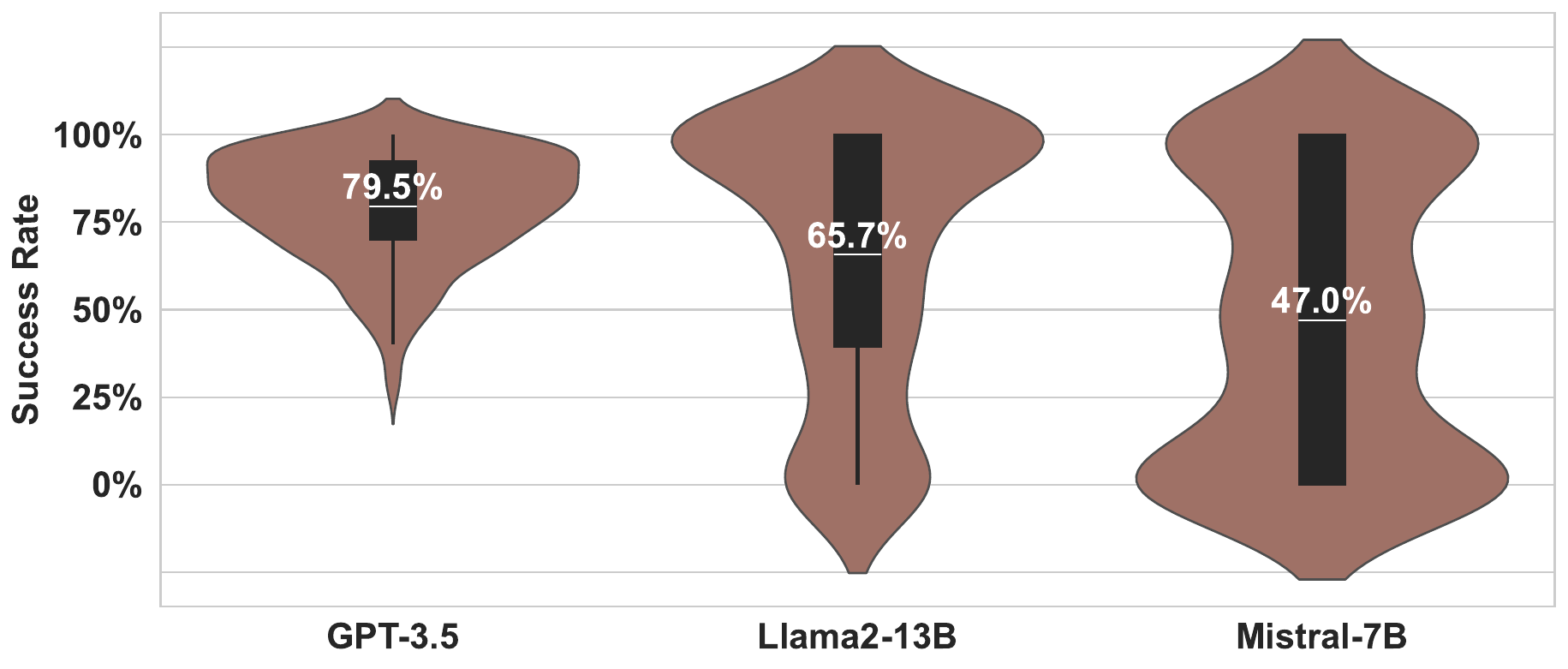}
    \caption{Distributions of each LLM's success rate on evaluating all driving scenarios. The means represent the central tendency.
    }    
    \label{fig:RQ0_success_rate_violin_diagram}
\end{figure}

\subsection{Research Questions (RQs)}\label{subsec:RQs}

Before evaluating the LLMs' robustness, we first check their success rates. From the violin plots in Figure~\ref{fig:RQ0_success_rate_violin_diagram}, we see the success rates 79.5\% for \gpt, 65.7\% for \llama and 47.0\% for \mistral. In addition, \gpt has the smallest variance. With a near 80\% success rate achieved by \gpt, we are confident that it is worth investigating the use of LLMs for determining the realism of driving scenarios. 
Based on this finding, we defined the following RQs. 
Table~\ref{tab:tabRQs} shows our RQs, various independent variables and their combinations, goals, and evaluation metrics, with which we study the robustness of LLMs from the following three perspectives.

\textbf{First}, we study the robustness of LLMs when ignoring the differences between various roads and weather conditions in RQ1. This allows us to study the robustness with respect to scenarios (individual and merged) and parameters (individual and merged). Consequently, our RQ1 is: How robust are LLMs in identifying realistic scenarios with respect to scenarios and parameters? RQ1 is divided into four sub-RQs: 1) \textbf{RQ1.a.} What is the overall robustness of LLMs in identifying realistic scenarios? This RQ studies LLMs' robustness without considering the differences in scenarios and parameters, providing a high-level view of each LLM's robustness; 2) \textbf{RQ1.b.} How robust are LLMs in identifying each realistic scenario? This RQ identifies scenarios for which LLMs' are the most and least robust in determining their realism; 3) \textbf{RQ1.c.} How robust are LLMs in identifying realistic scenarios with respect to each parameter? This RQ identifies parameters (e.g., position and velocity of an NPC vehicle) for which LLMs are the most and least robust; 4) \textbf{RQ1.d.} How robust are LLMs in identifying realistic scenarios with respect to each pair of scenarios and parameters? This RQ aims to identify the pairs of scenarios and parameters against which LLMs are the most and least robust.

\textbf{Second}, we study LLMs' robustness with respect to each road in RQ2 to study road characteristics that impact LLMs' robustness in identifying realistic scenarios. Thus, our RQ2 is: How robust are LLMs in identifying realistic scenarios on different roads? This RQ is further divided into four sub-RQs: 1) \textbf{RQ2.a.} What is the overall robustness of each LLM in identifying realistic scenarios on each road? This RQ identifies roads on which each LLM is the most and least robust in identifying the realism of scenarios; 2) \textbf{RQ2.b.} How robust are LLMs in identifying each realistic scenario on each road? This RQ identifies pairs of roads and scenarios against which LLM is the most and least robust; 3) \textbf{RQ2.c} How robust are LLMs in identifying realistic scenarios for each parameter on each road? This RQ helps us identify pairs of parameters and roads against which LLM is the least and the most robust in identifying the realism of scenarios; 4) \textbf{RQ2.d.} How robust are LLMs in identifying each realistic scenario on each road for each parameter? This RQ identifies the triplets of scenarios, parameters, and roads against which LLMs are the most and least robust. 

\textbf{Third}, we study LLMs' robustness with respect to each weather condition to study the weather conditions influencing LLMs' robustness in identifying realistic scenarios in RQ3. Our overall RQ3 is: How robust are LLMs in identifying realistic scenarios under various weather conditions? This RQ is answered with four sub-RQs: 1) \textbf{RQ3.a.} What is the overall robustness of LLMs for each weather condition? This RQ identifies weather conditions against which LLM is the least and the most robust in determining the realism of scenarios; 2) \textbf{RQ3.b.} How robust are LLMs in identifying each realistic scenario under each weather condition? This RQ identifies pairs of scenarios and weather conditions for which each LLM is the least and the most robust against; 3) \textbf{RQ3.c.} How robust are LLMs in identifying realistic scenarios for each pair of parameters and weather conditions? This RQ identifies pairs of parameters and weather conditions against which LLMs are the most and least robust in identifying realistic scenarios; 4) \textbf{RQ3.d.} How robust are LLMs in identifying each realistic scenario for each parameter under each weather condition? This RQ identifies the triplets of scenarios, parameters, and weather conditions against which LLMs are the most and least robust.  

\begin{table*}[]
\centering
\caption{Research questions, independent variables and metrics. \textit{Mer} denotes that data corresponding to an independent variable are merged during analyses, while \textit{Inv} means the otherwise. \textit{Sce}, \textit{Par} and \textit{WC} are abbreviations of scenario, parameter, and weather condition. Column \textit{Goal} describes based on which the robustness is studied. For instance, $Robustness \mid pair <sce, par>$ is about studying LLMs' robustness based on scenario and parameter pairs. In the formulas of the evaluation metrics, \textit{M1} and \textit{M2} denote any pair of \gpt, \llama, and \mistral.}
\label{tab:tabRQs}
\resizebox{\linewidth}{!}{

\begin{tabular}{cccc}
\toprule
\multirow{2}{*}{\textbf{RQ}}  & \textbf{Independent Variable} &  \multirow{2}{*}{\textbf{Goal}} & \multirow{2}{*}{\textbf{Evaluation Metrics (Section~\ref{subsec:metrics})}} \\ \cmidrule(r){2-2}
& \textbf{\begin{tabular}[c]{@{}c@{}}Sce\end{tabular}},
  \textbf{\begin{tabular}[c]{@{}c@{}}Par\end{tabular}},
  \textbf{\begin{tabular}[c]{@{}c@{}}Road\end{tabular}},
  \textbf{\begin{tabular}[c]{@{}c@{}}WC\end{tabular}} &  & \\
\midrule
1.a & Mer, Mer, Mer, Mer & Overall robustness &  $RS_{overall}=(\sum\nolimits_{sce}\sum\nolimits_{par}\sum\nolimits_{road}\sum\nolimits_{wc}RS_{<Org, Mut>}) / count(overall)$\\ \cmidrule(r){4-4}
1.b & Inv, Mer, Mer, Mer & Robustness $\mid$ sce & \textit{T-$5C_{all}$}, \textit{T-$5C_{M1,M2}$}, \textit{B-$5C_{all}$}, \textit{B-$5C_{M1,M2}$}, \textit{$IC_{M1,M2}$} \\ \cmidrule(r){4-4}
1.c & Mer, Inv, Mer, Mer & Robustness $\mid$ par & $RS_{par}=(\sum\nolimits_{sce}\sum\nolimits_{road}\sum\nolimits_{wc}RS_{<Org, Mut>})/count(par)$\\ \cmidrule(r){4-4}
1.d & Inv, Inv, Mer, Mer & Robustness $\mid$ pair <sce, par> & \textit{T-$5C_{all}$}, \textit{T-$5C_{M1,M2}$}, \textit{B-$5C_{all}$}, \textit{B-$5C_{M1,M2}$}, \textit{$IC_{M1,M2}$} \\ \midrule
2.a & Mer, Mer, Inv, Mer & Overall robustness $\mid$ road & $RS_{<overall,road>}=(\sum\nolimits_{sce}\sum\nolimits_{par}\sum\nolimits_{wc}RS_{<Org, Mut>})/count(<overall,road>)$\\ \cmidrule(r){4-4}
2.b & Inv, Mer, Inv, Mer & Robustness $\mid$ pair <sce, road> & \textit{T-$5C_{all}$}, \textit{T-$5C_{M1,M2}$}, \textit{B-$5C_{all}$}, \textit{B-$5C_{M1,M2}$}, \textit{$IC_{M1,M2}$} \\ \cmidrule(r){4-4}
2.c & Mer, Inv, Inv, Mer & Robustness $\mid$ pair <par, road> & $RS_{<par,road>}=(\sum\nolimits_{sce}\sum\nolimits_{wc}RS_{<Org, Mut>})/count(<par,road>)$\\ \cmidrule(r){4-4}
2.d & Inv, Inv, Inv, Mer & Robustness $\mid$ triplet <sce, par, road> & \textit{T-$5C_{all}$}, \textit{T-$5C_{M1,M2}$}, \textit{B-$5C_{all}$}, \textit{B-$5C_{M1,M2}$}, \textit{$IC_{M1,M2}$} \\ \midrule
3.a & Mer, Mer, Mer, Inv & Overall robustness $\mid$ wc & $RS_{<overall,wc>}=(\sum\nolimits_{sce}\sum\nolimits_{par}\sum\nolimits_{road}RS_{<Org, Mut>})/count(<overall,wc>)$\\ \cmidrule(r){4-4}
3.b & Inv, Mer, Mer, Inv & Robustness $\mid$ pair <sce, wc> & \textit{T-$5C_{all}$}, \textit{T-$5C_{M1,M2}$}, \textit{B-$5C_{all}$}, \textit{B-$5C_{M1,M2}$}, \textit{$IC_{M1,M2}$} \\ \cmidrule(r){4-4}
3.c & Mer, Inv, Mer, Inv & Robustness $\mid$ pair <par, wc> & $RS_{<par,wc>}=(\sum\nolimits_{sce}\sum\nolimits_{road}RS_{<Org, Mut>})/count(<par,wc>)$\\ \cmidrule(r){4-4}
3.d & Inv, Inv, Mer, Inv & Robustness $\mid$ triplet <sce, par, wc> & \textit{T-$5C_{all}$}, \textit{T-$5C_{M1,M2}$}, \textit{B-$5C_{all}$}, \textit{B-$5C_{M1,M2}$}, \textit{$IC_{M1,M2}$} \\
\bottomrule
\end{tabular}
}
\end{table*}

\subsection{Evaluation Metrics}\label{subsec:metrics}

\textbf{\textit{Robustness Score (\robustnessScore)}} 
is defined to measure each LLM's robustness in assessing the realism of driving scenarios. Since each LLM assesses each scenario $S$ 10 times, we use $\#Succ_{S}$ to denote the number of times an LLM successfully identifies $S$ as realistic, which ranges from 0-10. Next, for each original scenario $S_{Org}$ and one of its mutation scenario $S_{Mut}$ (e.g., position mutation $S_{{Mut}_{Pos}}$) pair, we define \robustnessScore as:
\begin{equation}
    RS_{<S_{Org}, S_{Mut}>}=\#AvgSuccess+\#Matched-\#Different,
\end{equation}
where $\#AvgSuccess$ is the average number of successes: $(\#Succ_{S_{Org}}+\#Succ_{S_{Mut}})/2$; 
$\#Matched$ is the number of successful results of $S_{Mut}$ that match successful results of $S_{Org}$: $\#Matched=min\{\#Succ_{S_{Org}}, \#Succ_{S_{Mut}}\}$; and $\#Different$ is the difference in the successful results of the LLM's evaluation on $S_{Org}$ and $S_{Mut}$: $abs(\#Succ_{S_{Org}} - \#Succ_{S_{Mut}})$. Therefore, the range of \robustnessScore is between -5 to 20. Here, \robustnessScore=20 indicates that the LLM successfully identifies both the original and mutated scenarios in all runs, hence maximum robustness, while \robustnessScore=-5 tells that the LLM is the least robust as the LLM identified the original (or mutated) scenario with a full extent of success, but failed for all runs on the mutated (original) scenario.
\robustnessScore=0, instead, tells that the LLM failed to identify both the original and mutated scenarios as realistic.

We then follow the goals defined in Table~\ref{tab:tabRQs} to calculate \robustnessScore for each RQ by merging or dividing scenarios into different categories and then averaging metric values in each category. 

\vspace{3pt}
\noindent \textbf{\textit{Top-5 Count (\topcount), Bottom-5 Count (\bottomcount), and Inconsistent Count (\inconsistentcount)}} are defined to compare the robustness of different LLMs in terms of assessing the scenario realism for each scenario. We first rank all the scenarios for each LLM according to the \robustnessScore achieved on each scenario. Next, we extract the top 5 and bottom 5 scenarios as two subsets, respectively: $\mathbb{S}_{top-5}^{M}$ and $\mathbb{S}_{bottom-5}^{M}$, where $M\in\{\gpt, \llama, \mistral\}$.

\topcount counts the number of identical scenarios in the top 5 scenarios of different LLMs. Concretely, we define \textit{T-$5C_{all}$} for all three LLMs as: 
\begin{equation}
    \textit{T-$5C_{all}$}=|\mathbb{S}_{top-5}^{\gpt}\cap\mathbb{S}_{top-5}^{\llama}\cap\mathbb{S}_{top-5}^{\mistral}|.
\end{equation}
Also, for each LLM pair M1 and M2, we define (\textit{T-$5C_{M1,M2}$}) as:
\begin{equation}
    \textit{T-$5C_{M1,M2}$}=|\mathbb{S}_{top-5}^{M1}\cap\mathbb{S}_{top-5}^{M2}|,
\end{equation}
where $M1\in\{\gpt, \llama, \mistral\}$ and $M2\in\{\gpt, \llama, \mistral\}\setminus\{M1\}$.

Similarly, \bottomcount counts the number of identical scenarios appearing in the bottom 5 scenarios of different LLMs. We calculate the \textit{B-$5C_{all}$} and \textit{B-$5C_{M1,M2}$} for all the three LLMs and LLM pairs.

\inconsistentcount measures the number of scenarios that fall into the top (bottom) 5 of M1 but the bottom (top) 5 of M2, i.e., appear inconsistently in the top 5 and bottom 5 of a scenario for the pair M1 and M2:
\begin{equation}
    \textit{$IC_{M1,M2}$}=|\mathbb{S}_{top-5}^{M1}\cap\mathbb{S}_{bottom-5}^{M2}| + |\mathbb{S}_{bottom-5}^{M1}\cap\mathbb{S}_{top-5}^{M2}|,
\end{equation}
where $M1\in\{\gpt, \llama, \mistral\}$ and $M2\in\{\gpt, \llama, \mistral\}\setminus\{M1\}$.

\section{Results and Analyses}\label{sec:results}

\subsection{RQ1: Overall Robustness of LLMs}\label{subsec:RQ1}

\begin{table*}[!htbp]
\centering
\caption{Robustness scores achieved by LLMs (overall and by parameter) for answering RQ1.a, RQ1.c, RQ2.a, RQ2.c, RQ3.a, and RQ3.c. The best robustness scores are highlighted in bold. \textit{all}, \textit{pos}, \textit{roa}, \textit{vel} and \textit{ang} denote parameters \textit{Overall}, \textit{Position}, \textit{Rotation}, \textit{Velocity}, and \textit{Angular Velocity}, respectively; \textit{R1-R4} represent Road1-Road4, and \textit{RD, RN, SD, and CN} denote Rainy Day, Rainy Night, Sunny Day, and Clear Night, respectively.}
\label{tab:tabRobustnessResults_a_c}
\resizebox{\linewidth}{!}{
\begin{tabular}{llrrrrr|rrrrr|rrrrr}
\toprule
\multicolumn{2}{c}{\textbf{RQ}} & \multicolumn{5}{c}{\textbf{GPT-3.5}} & \multicolumn{5}{c}{\textbf{Llama2-13B}} & \multicolumn{5}{c}{\textbf{Mistral-7B}} \\ \midrule
\textbf{} & \textbf{} & \multicolumn{1}{r}{\multirow{1}{*}{\textbf{$RS_{all}$}}} & \multicolumn{1}{r}{\multirow{1}{*}{\textbf{$RS_{pos}$}}} & \multicolumn{1}{r}{\multirow{1}{*}{\textbf{$RS_{roa}$}}} & \multicolumn{1}{r}{\multirow{1}{*}{\textbf{$RS_{vel}$}}} & \multicolumn{1}{r|}{\textbf{$RS_{ang}$}} & \multicolumn{1}{r}{\multirow{1}{*}{\textbf{$RS_{all}$}}} & \multicolumn{1}{r}{\multirow{1}{*}{\textbf{$RS_{pos}$}}} & \multicolumn{1}{r}{\multirow{1}{*}{\textbf{$RS_{roa}$}}} & \multicolumn{1}{r}{\multirow{1}{*}{\textbf{$RS_{vel}$}}} & \multicolumn{1}{r|}{\textbf{$RS_{ang}$}} & \multicolumn{1}{r}{\multirow{1}{*}{\textbf{$RS_{all}$}}} & \multicolumn{1}{r}{\multirow{1}{*}{\textbf{$RS_{pos}$}}} & \multicolumn{1}{r}{\multirow{1}{*}{\textbf{$RS_{roa}$}}} & \multicolumn{1}{r}{\multirow{1}{*}{\textbf{$RS_{vel}$}}} & \multicolumn{1}{r}{\textbf{$RS_{ang}$}} \\ \midrule
RQ1 & All & \textbf{12.59} & \textbf{12.88} & \textbf{12.23} & \textbf{12.38} & \textbf{12.87} & 9.48 & 8.85 & 9.22 & 9.89 & 9.97 & 5.60 & 5.05 & 6.11 & 4.79 & 6.44 \\ \midrule
\textbf{} & \textbf{} & \multicolumn{1}{r}{\multirow{1}{*}{\textbf{$RS_{<all,road>}$}}} & \multicolumn{1}{r}{\multirow{1}{*}{\textbf{$RS_{<pos,road>}$}}} & \multicolumn{1}{r}{\multirow{1}{*}{\textbf{$RS_{<roa,road>}$}}} & \multicolumn{1}{r}{\multirow{1}{*}{\textbf{$RS_{<vel,road>}$}}} & \multicolumn{1}{r|}{\textbf{$RS_{<ang,road>}$}} & \multicolumn{1}{r}{\multirow{1}{*}{\textbf{$RS_{<all,road>}$}}} & \multicolumn{1}{r}{\multirow{1}{*}{\textbf{$RS_{<pos,road>}$}}} & \multicolumn{1}{r}{\multirow{1}{*}{\textbf{$RS_{<roa,road>}$}}} & \multicolumn{1}{r}{\multirow{1}{*}{\textbf{$RS_{<vel,road>}$}}} & \multicolumn{1}{r|}{\textbf{$RS_{<ang,road>}$}} & \multicolumn{1}{r}{\multirow{1}{*}{\textbf{$RS_{<all,road>}$}}} & \multicolumn{1}{r}{\multirow{1}{*}{\textbf{$RS_{<pos,road>}$}}} & \multicolumn{1}{r}{\multirow{1}{*}{\textbf{$RS_{<roa,road>}$}}} & \multicolumn{1}{r}{\multirow{1}{*}{\textbf{$RS_{<vel,road>}$}}} & \multicolumn{1}{r}{\textbf{$RS_{<ang,road>}$}} \\ \midrule
\multirow{4}{*}{RQ2} & R1 & \textbf{12.64} & 12.70 & \textbf{12.23} & \textbf{13.08} & \textbf{12.55} & 11.89 & \textbf{13.36} & 12.00 & 10.47 & 11.72 & 5.89 & 4.72 & 9.69 & 4.22 & 4.92 \\
 & R2 & \textbf{12.20} & \textbf{12.34} & \textbf{12.39} & \textbf{11.55} & \textbf{12.50} & 6.12 & 4.06 & 6.28 & 6.08 & 8.06 & 3.74 & 3.88 & 0.78 & 3.81 & 6.48 \\
 & R3 & \textbf{12.38} & \textbf{12.78} & \textbf{11.67} & \textbf{12.55} & \textbf{12.50} & 6.46 & 4.50 & 6.41 & 8.59 & 6.33 & 5.89 & 5.38 & 6.33 & 4.61 & 7.23 \\
 & R4 & 13.15 & \textbf{13.70} & \textbf{12.61} & 12.36 & \textbf{13.92} & \textbf{13.47} & 13.48 & 12.20 & \textbf{14.44} & 13.77 & 6.88 & 6.23 & 7.66 & 6.53 & 7.11 \\ \midrule
 \textbf{} & \textbf{} & \multicolumn{1}{r}{\multirow{1}{*}{\textbf{$RS_{<all,wc>}$}}} & \multicolumn{1}{r}{\multirow{1}{*}{\textbf{$RS_{<pos,wc>}$}}} & \multicolumn{1}{r}{\multirow{1}{*}{\textbf{$RS_{<roa,wc>}$}}} & \multicolumn{1}{r}{\multirow{1}{*}{\textbf{$RS_{<vel,wc>}$}}} & \multicolumn{1}{r|}{\textbf{$RS_{<ang,wc>}$}} & \multicolumn{1}{r}{\multirow{1}{*}{\textbf{$RS_{<all,wc>}$}}} & \multicolumn{1}{r}{\multirow{1}{*}{\textbf{$RS_{<pos,wc>}$}}} & \multicolumn{1}{r}{\multirow{1}{*}{\textbf{$RS_{<roa,wc>}$}}} & \multicolumn{1}{r}{\multirow{1}{*}{\textbf{$RS_{<vel,wc>}$}}} & \multicolumn{1}{r|}{\textbf{$RS_{<ang,wc>}$}} & \multicolumn{1}{r}{\multirow{1}{*}{\textbf{$RS_{<all,wc>}$}}} & \multicolumn{1}{r}{\multirow{1}{*}{\textbf{$RS_{<pos,wc>}$}}} & \multicolumn{1}{r}{\multirow{1}{*}{\textbf{$RS_{<roa,wc>}$}}} & \multicolumn{1}{r}{\multirow{1}{*}{\textbf{$RS_{<vel,wc>}$}}} & \multicolumn{1}{r}{\textbf{$RS_{<ang,wc>}$}} \\ \midrule
\multirow{4}{*}{RQ3} & RD & \textbf{12.65} & \textbf{12.52} & \textbf{12.95} & \textbf{12.36} & \textbf{12.77} & 10.12 & 10.66 & 8.81 & 10.22 & 10.78 & 5.18 & 5.77 & 5.77 & 4.19 & 5.02 \\
 & RN & 12.81 & \textbf{13.34} & 11.12 & 13.41 & 13.38 & \textbf{13.55} & 11.89 & \textbf{13.94} & \textbf{14.06} & \textbf{14.33} & 3.63 & 2.81 & 4.38 & 2.89 & 4.45 \\
 & SD & \textbf{11.34} & \textbf{10.91} & \textbf{12.70} & \textbf{10.00} & \textbf{11.73} & 6.79 & 6.22 & 6.78 & 6.92 & 7.25 & 5.87 & 4.50 & 6.45 & 5.09 & 7.42 \\
 & CN & \textbf{13.56} & \textbf{14.77} & \textbf{12.12} & \textbf{13.77} & \textbf{13.59} & 7.47 & 6.64 & 7.36 & 8.38 & 7.52 & 7.71 & 7.12 & 7.86 & 7.00 & 8.86 \\ \bottomrule
\end{tabular}
}
\end{table*}

\begin{figure}[!htbp]
    \centering
    \includegraphics[width=0.65\linewidth]{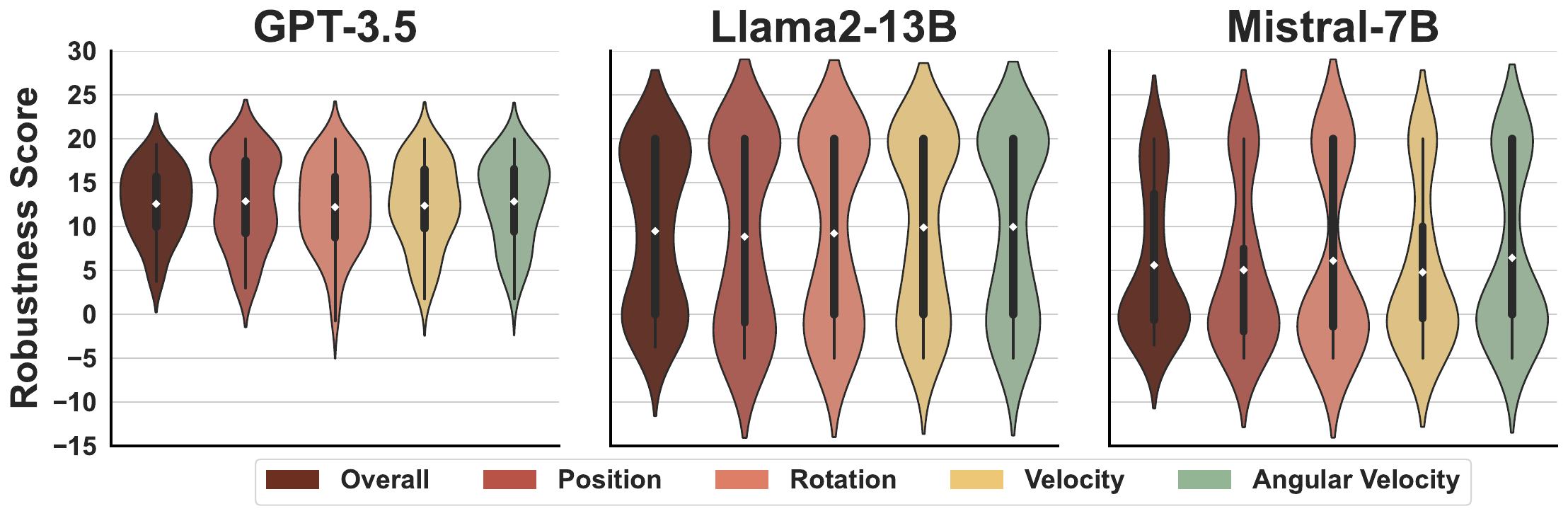}
    \caption{Distributions of robustness scores achieved by the LLMs (overall) for answering RQ1.a and RQ1.c. The mean represents the central tendency.}
    \label{fig:RQ1_a_c_robustness_violin_diagram}
\end{figure}

RQ1.a studies the overall LLMs' robustness without differentiating scenarios, parameters, roads, and weather conditions. Results are shown in the third row of Table~\ref{tab:tabRobustnessResults_a_c} in the \textit{$RS_{all}$} columns calculated with $RS_{overall}$. Overall, we can see that \gpt has the highest robustness score (i.e., $RS_{overall}=12.59$), followed by \llama (i.e., $RS_{overall}=9.48$), whereas \mistral has the lowest robustness score (i.e., $RS_{overall}=5.60$). Figure~\ref{fig:RQ1_a_c_robustness_violin_diagram} shows violin plots for the three models for \textit{Overall}. We observe that variances in robustness scores for \llama and \mistral are much higher compared to \gpt. Hence, we can conclude that \gpt demonstrates better robustness in accurately identifying realistic scenarios, exhibiting lower variance and thereby showing a higher level of trustworthiness in identifying realistic scenarios. 

In RQ1.b, we study the overall LLMs' robustness while differentiating scenarios. We use \textit{T-$5C_{all}$}, \textit{T-$5C_{M1,M2}$}, \textit{B-$5C_{all}$}, \textit{B-$5C_{M1,M2}$}, and $IC_{M1,M2}$ metrics as described in Section~\ref{subsec:metrics}. The results are summarized in Table~\ref{tab:tabRobustnessResults_RQ1_b_d_count} in the \textit{Overall} row. Generally, we have the following observations: 1) There is only one scenario that is identified as realistic by the three selected models in \textit{T-$5C_{all}$}, whereas there is no scenario identified as realistic with worst robustness by all three models in \textit{B-$5C_{all}$}. This suggests that specific characteristics of scenarios affect LLM's performance in correctly and robustly identifying scenarios as realistic; 2) We find scenarios commonly identified as realistic by two models in the same rank category. For instance, 6 scenarios (the highest number among all three pairs of models) ranked in \textit{T-$5C_{M1,M2}$} by \llama are also identified as realistic by \mistral with high robustness scores. This may suggest that \llama shares some abilities with \mistral to identify scenarios as realistic correctly with high robustness. Similarly, 5 scenarios are commonly identified by \gpt and \llama in \textit{T-$5C_{M1,M2}$} suggesting some similarity between the two models. 
3) We find that, generally, common scenarios (i.e., identical scenarios in the top 5 scenarios across two or more LLMs) are not identified in \textit{B-5C} as we can see that there is only one scenario that is identified as realistic with low robustness by \gpt and \mistral. 
Lastly, 4) the instances for \textit{IC} are also generally low with the highest $IC_{M1,M2}$ observed for the \llama and \mistral pair (i.e., 4).

These observations indicate that, in the context of identifying driving scenarios in terms of their realism, it is, in general, hard to achieve consensus among models, especially when more than two models are considered. This is because, firstly, driving scenarios contain rich information, which is more challenging to classify than many other tasks such as requirement classification~\cite{luo2022prcbert}. Secondly, the LLMs we experimented with are all general models, i.e., not made for ADS. 

RQ1.c analyzes LLMs' overall robustness by each parameter. Table~\ref{tab:tabRobustnessResults_a_c}, the second row (\textit{$RS_{pos}$}, \textit{$RS_{roa}$}, \textit{$RS_{vel}$}, and \textit{$RS_{ang}$} columns), shows the robustness scores (i.e., $RS_{par}$) of the four parameters for the three models. Generally, we have the same findings as RQ1.a, i.e., \gpt has the highest $RS_{par}$ for all four parameters, followed by slightly lower $RS_{par}$ for \llama. Finally, \mistral has the lowest $RS_{par}$. Figure~\ref{fig:RQ1_a_c_robustness_violin_diagram} shows violin plots for the three models for each parameter. We see that \gpt has the lowest variance in robustness scores across all parameters.

In RQ1.d, we study the overall robustness of LLMs while considering each pair of scenarios and parameters. Table~\ref{tab:tabRobustnessResults_RQ1_b_d_count}, rows 3-6 summarize the results for the four parameters. Generally, the \llama and \mistral pair, for all the four parameters have the highest number of \textit{T-5C}, \textit{B-5C}, and \textit{IC} with one exception (i.e., for \textit{Angular Velocity}, it is second best after the \gpt and \llama pair tied with the rest of the model pairs). These results suggest that \llama and \mistral share some abilities that make them consistently identify the same scenarios as realistic. On the other hand, the \llama and \mistral pair also share the highest number of inconsistencies.

\finding{
\textbf{RQ1:} \gpt demonstrate the best robustness in identifying realistic scenarios with lower variance followed by \llama and \mistral. However, \llama and \mistral have large variances in their robustness scores. Moreover, robustness is also specific to each scenario indicating that the specific characteristics of scenarios, indeed, impact the robustness of LLMs. Finally, \llama and \mistral appear to have similar abilities to identify realistic scenarios with high robustness on all the parameters.  
}

\begin{table}[]
\centering
\caption{Top-5 Count (\textit{T-5C}), Bottom-5 Count (\textit{B-5C}), and Inconsistent Count (\textit{IC}) (overall) for answering RQ1.b and RQ1.d. 
\textit{all} denote \textit{T-$\bm{5C_{all}}$}/\textit{B-$\bm{5C_{all}}$}. \textit{G,L}, \textit{G,M}, and \textit{L,M} represent \textit{T-$\bm{5C_{M1,M2}}$}/\textit{B-$\bm{5C_{M1,M2}}$}/\textit{$\bm{IC_{M1,M2}}$} with the model pairs (i.e., \textit{M1}, \textit{M2}) of \gpt and \llama, \gpt and \mistral, and \llama and \mistral, respectively.
}
\label{tab:tabRobustnessResults_RQ1_b_d_count}
\vspace{5pt}
\resizebox{0.7\linewidth}{!}{
\begin{tabular}{lrrrr|rrrr|rrr}
\toprule
\multirow{2}{*}{\textbf{Parameter}} & \multicolumn{4}{r|}{\textbf{\textit{T-5C}}} & \multicolumn{4}{r|}{\textbf{\textit{B-5C}}} & \multicolumn{3}{r}{\textbf{\textit{IC}}} \\ \cmidrule(r){2-12} 
 & \textbf{\textit{all}} & \textbf{\textit{G,L}} & \textbf{\textit{G,M}} & \textbf{\textit{L,M}} & \textbf{\textit{all}} & \textbf{\textit{G,L}} & \textbf{\textit{G,M}} & \textbf{\textit{L,M}} & \textbf{\textit{G,L}} & \textbf{\textit{G,M}} & \textbf{\textit{L,M}} \\ \midrule
Overall & 1 & 5 & 2 & \textbf{6} & 0 & 0 & \textbf{1} & 0 & 1 & 1 & \textbf{4} \\
Position & 10 & 13 & 17 & \textbf{26} & 0 & 3 & 1 & \textbf{5} & 7 & 9 & \textbf{17} \\
Rotation & 3 & 9 & 5 & \textbf{12} & 0 & 1 & 1 & \textbf{4} & 5 & 8 & \textbf{11} \\
Velocity & 7 & 12 & 7 & \textbf{21} & 1 & 2 & 3 & \textbf{4} & 7 & 4 & \textbf{14} \\
Angular Velocity & 7 & \textbf{15} & 10 & 14 & 1 & 1 & \textbf{3} & 1 & 4 & 7 & \textbf{10} \\ \bottomrule
\end{tabular}
}
\end{table}

\subsection{RQ2: LLMs Robustness by Roads} \label{subsec:RQ2}

\begin{figure}[!htbp]
    \centering
    \includegraphics[width=0.7\linewidth]{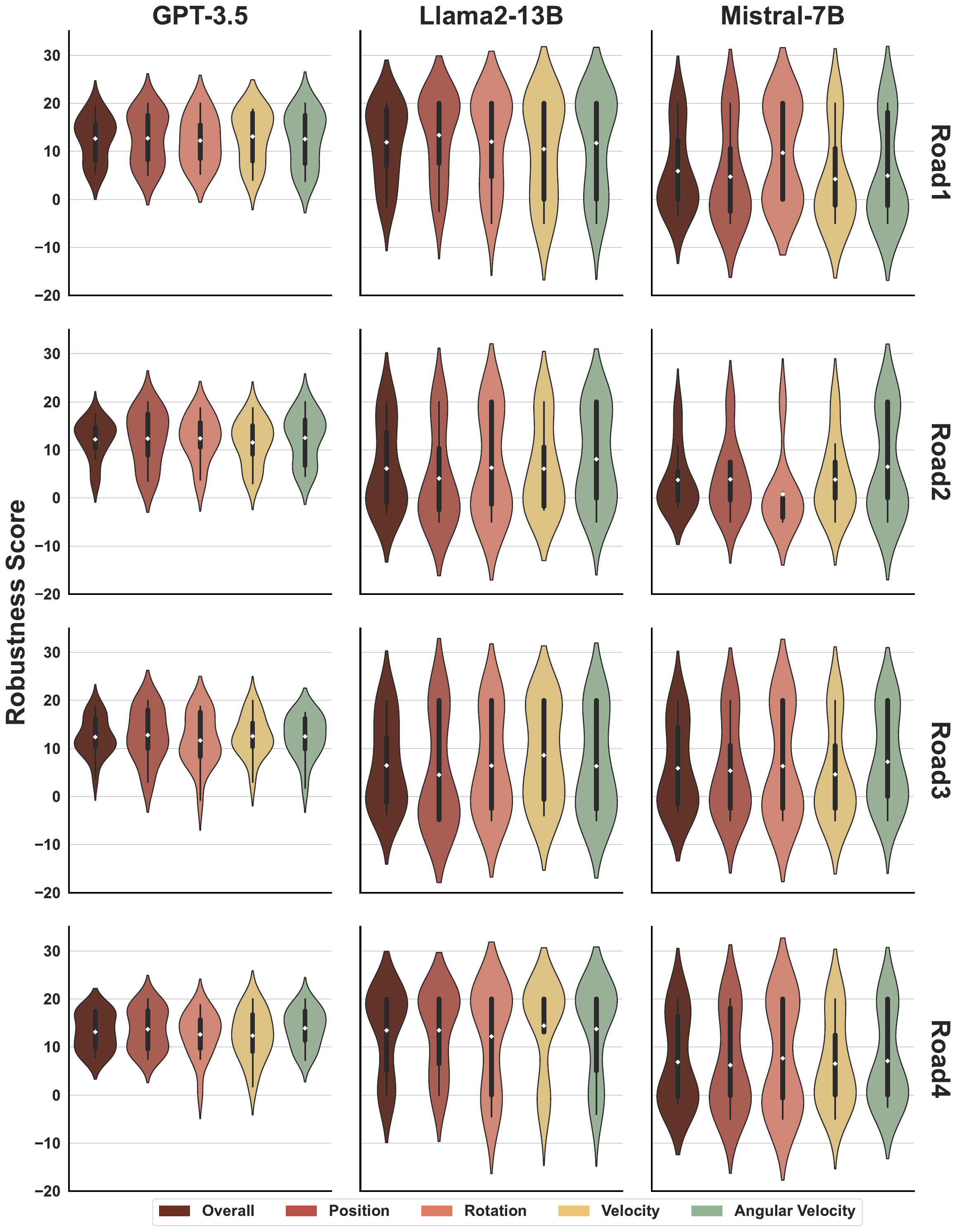}
    \caption{Distributions of robustness scores achieved by the LLMs (by road) for answering RQ2.a and RQ2.c. The mean represents the central tendency.}
    \label{fig:RQ2_a_c_robustness_violin_diagram}
\end{figure}

RQ2.a studies the overall LLMs' robustness for each road (i.e., $RS_{<overall,road>}$). Results are shown in Table~\ref{tab:tabRobustnessResults_a_c}. For R1-R3 with \textit{Overall}, \gpt has the highest robustness scores (i.e., 12.64, 12.20, and 12.38), whereas for R4, \llama has the highest robustness score, i.e., 13.47. \mistral is the worst for all roads. Our results indicate that road characteristics, indeed, impact the LLMs' robustness in identifying realistic driving scenarios. In addition, from Figure~\ref{fig:RQ2_a_c_robustness_violin_diagram}, we can further observe that \gpt has the lowest variance for all roads. Moreover, for R4, all models (especially \gpt) achieved slightly less variance than for the other roads. 
After investigating the road structure of R4 and the scenarios generated on R4, we find that it is more difficult for the scenario generation strategy (i.e., DeepQTest) to generate realistic scenarios on R4 than on the other roads because the road structure of R4 is more complex than the others~\cite{lu2023deepqtest}. 
Thus, we conjecture that the generated realistic scenarios for R4 are complex realistic scenarios, and LLMs are good at comprehending complex scenarios.

RQ2.b studies the LLMs' robustness for each road and scenario pair with the \textit{T-$5C_{all}$}, \textit{T-$5C_{M1,M2}$}, \textit{B-$5C_{all}$}, \textit{B-$5C_{M1,M2}$}, and $IC_{M1,M2}$ metrics as described in Section~\ref{subsec:metrics}. The results are shown in Table~\ref{tab:tabRobustnessResults_RQ2_b_d_count} (see rows \textit{Overall} for each road). 
Generally, we find that for \textit{T-5C}, the pair \llama and \mistral has the highest count for R1-R3, whereas for R4, the pair \gpt and \llama has the highest count, but only one more than the \llama and \mistral pair (i.e., a minor difference). These results indicate that for R1-R3, \llama and \mistral share the abilities to identify realistic scenarios robustly, whereas the same holds for R4, \gpt and \llama. For \textit{B-5C}, for \llama and \mistral pair, we observed the highest number of counts for R1 and R2, whereas for R3 and R4, the pair has the second highest number of counts. In general, for both, \textit{T-5C} and \textit{B-5C}, the highest count is quite small (i.e., the highest one is 4). This indicates that there are some common patterns (e.g., similarity between the \llama and \mistral pair), but those appear to be weak. For \textit{IC}, the \llama and \mistral pair has the highest count for R1, R3, and R4, whereas the second highest for R2 (which is one less than the highest). This finding indicates that the pair \llama and \mistral also tends to give inconsistent results in terms of identifying realistic scenarios. A possible reason for this could be that both \llama and \mistral show low confidence when predicting scenarios' realism, producing higher inconsistencies among them.

RQ2.c studies the robustness ($RS_{<par,road>}$) of the LLMs for each road and parameter pair. The results are shown in Table~\ref{tab:tabRobustnessResults_a_c}, under the RQ2 row under the $RS_{<pos,road>}$, $RS_{<roa,road>}$, $RS_{<vel,road>}$, and $RS_{<ang,road>}$ columns. For R1-R4, in general, \gpt has the highest robustness for all the parameters except \textit{Position} on R1 and \textit{Velocity} on R4, where \llama has a better robustness. In addition, from the violin plots in Figure~\ref{fig:RQ2_a_c_robustness_violin_diagram}, we observe that for all roads and all parameters, \gpt still has low variance in robustness scores, indicating \gpt producing reliable results. Based on these results, we can conclude that \gpt, for most parameters on all roads, robustly identifies realistic scenarios. Although in two specific cases \llama's robustness was slightly higher than \gpt's, in R2 and R3 roads, \llama's values plummeted. This suggests that \llama is robust only under certain circumstances, e.g., the R1 and \textit{Position} pair. 


\begin{table}[]
\centering
\caption{Top-5 Count (\textit{T-5C}), Bottom-5 Count (\textit{B-5C}), and Inconsistent Count (\textit{IC}) (by road) for answering RQ2.b and RQ2.d. Columns \textit{all} denote \textit{T-$\bm{5C_{all}}$}/\textit{B-$\bm{5C_{all}}$}. Columns \textit{G,L}, \textit{G,M}, and \textit{L,M} represent \textit{T-$\bm{5C_{M1,M2}}$}/\textit{B-$\bm{5C_{M1,M2}}$}/\textit{$\bm{IC_{M1,M2}}$} with the model pairs (i.e., \textit{M1}, \textit{M2}) of \gpt and \llama, \gpt and \mistral, and \llama and \mistral, respectively. \textit{R1-R4} represent Road1-Road4.
}
\vspace{5pt}
\resizebox{0.8\linewidth}{!}{
\begin{tabular}{llrrrr|rrrr|rrr}
\toprule
\multirow{2}{*}{\textbf{Road}} & \multirow{2}{*}{\textbf{Parameter}} & \multicolumn{4}{r|}{\textbf{\textit{T-5C}}} & \multicolumn{4}{r|}{\textbf{\textit{B-5C}}} & \multicolumn{3}{r}{\textbf{\textit{IC}}} \\ \cmidrule(r){3-13} 
 & & \textbf{\textit{all}} & \textbf{\textit{G,L}} & \textbf{\textit{G,M}} & \textbf{\textit{L,M}} & \textbf{\textit{all}} & \textbf{\textit{G,L}} & \textbf{\textit{G,M}} & \textbf{\textit{L,M}} & \textbf{\textit{G,L}} & \textbf{\textit{G,M}} & \textbf{\textit{L,M}} \\ \midrule
\multirow{5}{*}{R1} & Overall & 1 & 2 & 2 & \textbf{3} & 1 & 3 & 2 & \textbf{4} & 5 & 6 & \textbf{8} \\
 & Position & 6 & 7 & 6 & \textbf{14} & 5 & 6 & 5 & \textbf{12} & 13 & 10 & \textbf{26} \\
 & Rotation & 6 & 6 & 7 & \textbf{14} & 2 & 2 & 6 & \textbf{7} & 8 & 13 & \textbf{21} \\
 & Velocity & 7 & 7 & 8 & \textbf{14} & 3 & 3 & 5 & \textbf{9} & 10 & 13 & \textbf{23} \\
 & Angular Velocity & 5 & 7 & 5 & \textbf{13} & 2 & 5 & 2 & \textbf{11} & 12 & 10 & \textbf{24} \\ \midrule
\multirow{5}{*}{R2} & Overall & 1 & \textbf{3} & 2 & \textbf{3} & 1 & 2 & \textbf{3} & \textbf{3} & 3 & \textbf{5} & 4 \\
 & Position & 6 & 6 & 7 & \textbf{8} & 5 & 6 & 5 & \textbf{9} & 5 & 9 & \textbf{11} \\
 & Rotation & 3 & 4 & 5 & \textbf{8} & 3 & 3 & 4 & \textbf{9} & 6 & 10 & \textbf{16} \\
 & Velocity & 2 & 2 & 5 & \textbf{10} & 4 & 4 & 5 & \textbf{11} & 8 & 7 & \textbf{18} \\
 & Angular Velocity & 5 & 5 & 6 & \textbf{13} & 3 & 3 & \textbf{7} & \textbf{7} & 9 & 13 & \textbf{20} \\ \midrule
\multirow{5}{*}{R3} & Overall & 3 & \textbf{4} & 3 & \textbf{4} & 1 & 2 & \textbf{3} & 2 & 4 & 3 & \textbf{6} \\
 & Position & 6 & 6 & 7 & \textbf{10} & 3 & 4 & \textbf{6} & \textbf{6} & 6 & 12 & \textbf{14} \\
 & Rotation & 3 & 4 & 5 & \textbf{10} & 1 & 3 & 2 & \textbf{5} & 7 & 9 & \textbf{15} \\
 & Velocity & 3 & 4 & 4 & \textbf{8} & 1 & 3 & \textbf{4} & \textbf{4} & 7 & 7 & \textbf{13} \\
 & Angular Velocity & 4 & 5 & 7 & \textbf{10} & 2 & 4 & 2 & \textbf{6} & 7 & 11 & \textbf{17} \\ \midrule
\multirow{5}{*}{R4} & Overall & 2 & \textbf{5} & 3 & 4 & 1 & \textbf{2} & \textbf{2} & 1 & 6 & 3 & \textbf{8} \\
 & Position & 5 & 7 & 5 & \textbf{13} & 5 & 6 & 5 & \textbf{9} & 13 & 9 & \textbf{22} \\
 & Rotation & 3 & 4 & 4 & \textbf{12} & 1 & 2 & \textbf{4} & 3 & 7 & 8 & \textbf{14} \\
 & Velocity & 4 & 5 & 4 & \textbf{13} & 4 & 5 & 4 & \textbf{9} & 10 & 7 & \textbf{22} \\
 & Angular Velocity & 8 & 8 & 8 & \textbf{15} & 4 & 5 & 4 & \textbf{11} & 13 & 10 & \textbf{26} \\ \bottomrule
\end{tabular}
\label{tab:tabRobustnessResults_RQ2_b_d_count}
}
\end{table}

RQ2.d studies the LLMs' robustness concerning each triplet of scenario, parameter, and road using the T-$5C_{all}$, T-$5C_{M1,M2}$, B-$5C_{all}$, B-$5C_{M1,M2}$, and $IC_{M1,M2}$ metrics from Section \ref{subsec:metrics}. Table~\ref{tab:tabRobustnessResults_RQ2_b_d_count} summarizes the results (see \textit{Position}, \textit{Rotation}, \textit{Velocity}, and \textit{Angular Velocity} rows for all roads). In general, we find that for \textit{T-5C}, for all parameters, \llama and \mistral identify the highest numbers of common scenarios. A similar pattern is observed for \textit{B-5C}, i.e., \llama and \mistral have the highest common number of scenarios (except for \textit{Rotation} with R4, where it is second best). These results, once again, indicate similar abilities of \llama and \mistral to common scenarios with high robustness or low robustness. With respect to \textit{Inconsistent}, the \llama and \mistral pair has the highest number of counts. 

\finding{
\textbf{RQ2:} Road characteristics, indeed, affect LLMs robustness in identifying realistic scenarios. For most roads, \gpt still has the highest robustness with low variances, whereas for one road \llama has the highest robustness (although the difference with \gpt is low). Moreover, in roads R2 and R3, the $RS$ of \llama are very low when compared to \gpt. \mistral has the worst robustness on all the roads. The pair \llama and \mistral tends to show similarities in identifying realistic scenarios with high robustness on all the roads.  
}

\begin{figure}[!htbp]
    \centering
    \includegraphics[width=0.7\linewidth]{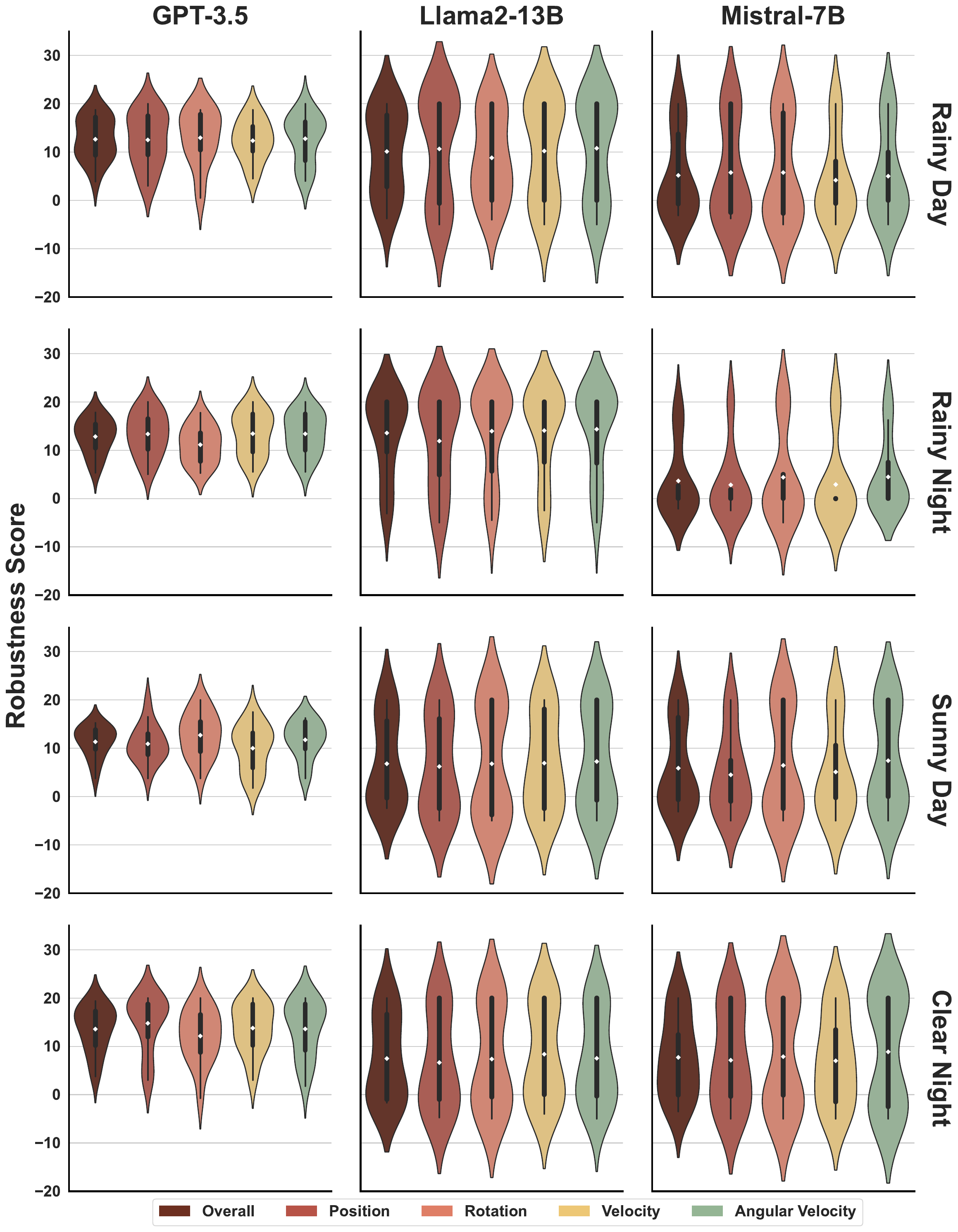}
    \caption{Distributions of robustness scores achieved by the LLMs (by weather condition) for answering RQ3.a and RQ3.c. The mean represents the central tendency.}
    \label{fig:RQ3_a_c_robustness_violin_diagram}
\end{figure}

\subsection{RQ3: LLMs Robustness by Weather Conditions} \label{subsec:RQ3}
We study the effect of weather conditions on the LLMs' robustness. Particularly, RQ3.a studies the overall LLMs' robustness for each weather condition. Results are shown in the RQ3 row, \textit{Overall} of Table~\ref{tab:tabRobustnessResults_a_c}. For RD, SD, and CN, \gpt performs the most robust (12.65 for RD, 11.34 for SD and 13.56 for CN). For RN, \llama outperforms both \gpt and \mistral, though the difference between \llama and \gpt is very close (13.55 vs. 12.81). Under all four weather conditions, \gpt and \llama largely outperform \mistral; in other words, \mistral is consistently the worst model. Furthermore, we can observe that, under RD and RN, the performance of \gpt and \llama is close, but \llama performs very poorly for SD and CN as compared to \gpt. These results indicate that the weather conditions do impact LLMs' ability to identify realistic scenarios robustly and the level of the impact differs; \gpt is consistently robust and \mistral is consistently not robust under all four weather conditions. Interestingly, \llama is much more robust on the rainy day and night, but less robust on the sunny day and clear night.

\begin{table}[]
\centering
\caption{Top-5 Count (\textit{T-5C}), Bottom-5 Count (\textit{B-5C}), and Inconsistent Count (\textit{IC}) (by weather condition) for answering RQ3.b and RQ3.d. Columns \textit{all} denote \textit{T-$\bm{5C_{all}}$}/\textit{B-$\bm{5C_{all}}$}. Columns \textit{G,L}, \textit{G,M} and \textit{L,M} represent \textit{T-$\bm{5C_{M1,M2}}$}/\textit{B-$\bm{5C_{M1,M2}}$}/\textit{$\bm{IC_{M1,M2}}$} with the model pairs (i.e., \textit{M1}, \textit{M2}) of \gpt and \llama, \gpt and \mistral, and \llama and \mistral, respectively. \textit{RD, RN, SD, and CN} denote Rainy Day, Rainy Night, Sunny Day, and Clear Night.
}
\label{tab:tabRobustnessResults_RQ3_b_d_count}
\vspace{5pt}
\resizebox{0.8\linewidth}{!}{
\begin{tabular}{llrrrr|rrrr|rrr}
\toprule
\multirow{2}{*}{\textbf{Weather}} & \multirow{2}{*}{\textbf{Parameter}} & \multicolumn{4}{r|}{\textbf{\textit{T-5C}}} & \multicolumn{4}{r|}{\textbf{\textit{B-5C}}} & \multicolumn{3}{r}{\textbf{\textit{IC}}} \\ \cmidrule(r){3-13} 
 & & \textbf{\textit{all}} & \textbf{\textit{G,L}} & \textbf{\textit{G,M}} & \textbf{\textit{L,M}} & \textbf{\textit{all}} & \textbf{\textit{G,L}} & \textbf{\textit{G,M}} & \textbf{\textit{L,M}} & \textbf{\textit{G,L}} & \textbf{\textit{G,M}} & \textbf{\textit{L,M}} \\ \midrule
\multirow{5}{*}{RD} & Overall & 2 & \textbf{5} & 3 & 3 & 1 & \textbf{3} & \textbf{3} & \textbf{3} & 3 & 4 & \textbf{9} \\
 & Position & 3 & 7 & 3 & \textbf{10} & 4 & 5 & 4 & \textbf{9} & 12 & 8 & \textbf{19} \\
 & Rotation & 4 & 6 & 5 & \textbf{8} & 0 & \textbf{2} & 1 & \textbf{2} & 7 & 8 & \textbf{9} \\
 & Velocity & 6 & 8 & 6 & \textbf{10} & 2 & 3 & 4 & \textbf{5} & 6 & 9 & \textbf{17} \\
 & Angular Velocity & 4 & 6 & 5 & \textbf{11} & 0 & 2 & \textbf{3} & \textbf{3} & 7 & 10 & \textbf{17} \\ \midrule
\multirow{5}{*}{RN} & Overall & 4 & 4 & 5 & \textbf{11} & 3 & 3 & 6 & \textbf{7} & 8 & 8 & \textbf{16} \\
 & Position & 6 & 6 & 7 & \textbf{14} & 3 & 3 & 6 & \textbf{8} & 10 & 13 & \textbf{22} \\
 & Rotation & 6 & 6 & 6 & \textbf{16} & 6 & 6 & 6 & \textbf{16} & 12 & 12 & \textbf{32} \\
 & Velocity & 6 & 6 & 6 & \textbf{16} & 8 & 8 & 8 & \textbf{16} & 14 & 14 & \textbf{32} \\
 & Angular Velocity & 10 & 10 & 10 & \textbf{16} & 5 & 5 & 5 & \textbf{16} & 15 & 15 & \textbf{32} \\ \midrule
\multirow{5}{*}{SD} & Overall & 1 & \textbf{3} & 1 & \textbf{3} & 1 & \textbf{2} & \textbf{2} & 1 & 2 & \textbf{5} & \textbf{5} \\
 & Position & 2 & 3 & 4 & \textbf{5} & 4 & 5 & \textbf{7} & 6 & 6 & 11 & \textbf{14} \\
 & Rotation & 5 & 6 & 7 & \textbf{9} & 3 & \textbf{5} & 3 & \textbf{5} & 4 & 9 & \textbf{12} \\
 & Velocity & 4 & 4 & 5 & \textbf{10} & 2 & 2 & 4 & \textbf{6} & 8 & 7 & \textbf{16} \\
 & Angular Velocity & 7 & 8 & 8 & \textbf{10} & 2 & 3 & 3 & \textbf{5} & 6 & 9 & \textbf{15} \\ \midrule
\multirow{5}{*}{CN} & Overall & 2 & 3 & 2 & \textbf{6} & 1 & 2 & 2 & \textbf{3} & \textbf{4} & 3 & \textbf{4} \\
 & Position & 9 & 10 & \textbf{11} & \textbf{11} & 2 & 2 & 4 & \textbf{5} & 7 & 12 & \textbf{11} \\
 & Rotation & 4 & 4 & 6 & \textbf{13} & 3 & 3 & 6 & \textbf{9} & 8 & 12 & \textbf{22} \\
 & Velocity & 5 & 7 & 5 & \textbf{10} & 2 & 3 & 3 & \textbf{5} & \textbf{11} & 8 & 9 \\
 & Angular Velocity & 6 & 7 & 8 & \textbf{11} & 2 & 3 & 3 & \textbf{4} & 9 & 12 & \textbf{13} \\ \bottomrule
\end{tabular}
}
\end{table}

From Figure~\ref{fig:RQ3_a_c_robustness_violin_diagram}, one can clearly observe that \gpt achieves the lowest variances in the robustness scores under all weather conditions. Both \llama and \mistral show large variances under all weather conditions. Though for RD, \llama achieves higher robustness scores, the variance of the distribution is much larger than \gpt's, hence \gpt is recommended.  

RQ3.b studies the robustness with respect to each weather condition and scenario pair (see results in the \textit{Overall} rows of Table~\ref{tab:tabRobustnessResults_RQ3_b_d_count}). Under RD, for \textit{T-5C}, the \gpt and \llama pair (\textit{G,L}) achieve the highest number of common scenarios (i.e., 5); for \textit{B-5C}, three pairs are equal in sharing 3 common scenarios. Under RN and CN, the pair of \llama and \mistral shares the highest numbers of common scenarios in both \textit{T-5C} and \textit{B-5C}, implying that these two models agree with each other the most under rainy and clear nights. For SD, we do not see a clear champion. When looking at \textit{IC}, under all four weather conditions, the pair of \llama and \mistral has the highest numbers, implying that they disagree the most. One plausible reason could be that both \llama and \mistral demonstrate low confidence in identifying realistic scenarios, introducing more non-determinism in the evaluation results.


RQ3.c studies robustness across weather conditions and parameters. Results are shown in the RQ3 row of Table~\ref{tab:tabRobustnessResults_a_c}. Under RD, SD and CN, consistently across all parameters, \gpt has the best robustness. Under RN, \gpt performs the best for \textit{Position}, but \llama performs the best for the other three parameters. \mistral performs consistently poorly under all weather conditions for all parameters. In general, \llama and \gpt achieve comparable results under RD and RN, but \llama performs much worse under SD and CN. 
These results are quite consistent with the observations from RQ3.a. From Figure~\ref{fig:RQ3_a_c_robustness_violin_diagram}, we can observe the low variance in robustness scores for \gpt. However, it is hard to observe any interaction effects between the weather conditions and parameters.   

RQ3.d studies the robustness of the LLMs by each triplet of scenarios, parameters, and weather conditions. Results are shown in Table~\ref{tab:tabRobustnessResults_RQ3_b_d_count} (see \textit{Position}, \textit{Rotation}, \textit{Velocity}, and \textit{Angular Velocity} under each weather condition). For \textit{T-5C}, we observe that \llama and \mistral share the highest numbers of common scenarios for all the four weather conditions with the exception of CN with Position, where the \gpt and \mistral pair has the same number of common scenarios (i.e., 11) as the \llama and \mistral pair. Similarly, for \textit{B-5C}, under RN and CN, the \llama and \mistral pair has the highest numbers of common scenarios across all parameters; under the other two weather conditions, the \llama and \mistral pair also share more common scenarios in most cases, but the differences with the other pairs are not so prominent. For \textit{IC}, the \llama and \mistral pair disagrees the most.

\finding{
\textbf{RQ3:} \gpt is the best under three weather conditions and slightly under-performed \llama under the rainy night, but with much smaller variances. \llama performed much better under the rainy day and rainy night conditions than under the sunny day and clear night conditions. \mistral performed poorly consistently under all weather conditions. These observations are also obtained when looking at each parameter.
}

\section{Threats to Validity}\label{sec:threat}
\vspace{3pt}
\textbf{\textit{Conclusion Validity}}. LLMs are stochastic, which means that they produce different results when running the same prompt (even when the temperature parameter value is 0~\cite{ouyang2023llm}). We mitigated this threat by running each prompt 10 times and by considering this stochasticity in the evaluation metrics (Section~\ref{subsec:metrics}). 

\vspace{3pt}
\noindent\textbf{\textit{Construct Validity}}. LLMs have a temperature parameter that must be set up, which we set to 0 because our application does not require diversity and we were interested in checking realism deterministically. 
Each LLM has its own specific default prompt format, e.g., the \textit{<<SYS>> {{system prompt}} <</SYS>>} for \llama. Not following such a format affect the LLM's evaluation capabilities. To mitigate this threat, we carefully followed the documentation of each LLM to correctly use the default format with each prompt, while keeping the main contents of the prompt same for all LLMs to enable fair comparison among the selected LLMs.

\vspace{3pt}
\noindent\textbf{\textit{External Validity}}. First, we only selected 576 realistic scenarios with one ego vehicle and one NPC vehicle, which does not guarantee that the results will be generalized to other cases, e.g., realistic scenarios with more NPC vehicles or unrealistic scenarios. Given the input limitations of the LLMs, the length of the prompts would have increased too much (i.e., exceeding the maximum token size) if more NPC vehicles had been provided, which could have resulted in the LLMs failing to output valid answers. In the future, we will consider more LLMs supporting larger numbers of tokens and different prompt templates to mitigate this threat. In addition, we will systematically involve unrealistic scenarios to further study the LLM's performance in assessing unrealistic scenarios, to make our findings more generalizable. Another external validity threat is that we only used scenarios that were generated by \deepscenario. Our implementation followed a decoding strategy to easily integrate other scenarios with other formats, which will be addressed in future evaluations to further generalize our findings.

\section{Conclusion and Future Work}\label{sec:conclusion}
Testing autonomous driving systems (ADSs) focuses on generating driving scenarios. The realism of these scenarios plays a crucial role in assessing the dependability of ADSs. To this end, we empirically evaluated the robustness of three state-of-the-art large-language models (LLMs) in determining the realism of driving scenarios. For the empirical evaluation, we selected a total of 64 realistic scenarios from \deepscenario and generated additional 512 realistic scenarios by mutating the original scenarios’ parameters (e.g., velocity) with minor changes. We evaluated LLMs’ robustness from the perspectives of roads, weather conditions, scenarios, and scenario parameters individually and their combinations.

The results of our evaluation provide some interesting insights that are useful for practitioners designing ADS test generators. For instance, ADS testers can optimize the test generation process to generate realistic driving scenarios by selecting the LLM that is the most robust in recognizing realistic scenarios and then integrating it into the test generator to assess the realism of the generated driving scenarios.
The most interesting, although not surprising, finding was that \gpt is the most robust LLM for identifying realistic scenarios, showing the highest robustness scores with the lowest variances. Although, for some roads and weather conditions, \llama showed higher robustness scores, the performance of this model significantly degraded under other conditions. In contrast, \gpt's average robustness scores remained high for any kind of weather and road. Lastly, overall, \mistral showed the worst robustness among the three LLMs, and therefore, its use is not recommended for this specific task. Another interesting finding was that both road characteristics and weather conditions affected the LLMs' robustness when identifying realistic scenarios. For instance, \llama seems good at identifying realistic scenarios for roads R1 and R4, but bad on roads R2 and R3, and performs well under rainy night weather conditions. These results indicate that it might be worth selecting different LLMs according to the characteristics of road structures, weather conditions, and other environmental parameters. However, we need a dedicated and extensive experiment studying various environmental conditions in many scenarios to devise such an LLM selection strategy systematically.     


There are different avenues we would like to explore in the near future. Firstly, we would like to experiment with other models, such as \textit{GPT-4}; since \gpt showed the highest robustness, our intuition is that \textit{GPT-4} can even outperform \gpt. 
Secondly, our evaluation was limited to 1 NPC vehicle due to the problems with the token limitation. To solve this issue, we would like to explore alternative strategies to use more than 1 NCP (e.g., reducing the timestamps, or invoking several times the LLMs and merging all the different outcomes). 
Besides, we intend to include unrealistic driving scenarios in our future empirical study. To do so, we plan to design an unrealistic scenario generation method and evaluate the generated unrealistic scenarios to further enhance the generalizability of our findings. 
Lastly, the LLMs we experimented with are all general models with no particular expertise in the field of ADSs. Thus, in future studies, we would like to study whether strategies like fine-tuning could help increase the robustness of LLMs when identifying the realism of ADSs' scenarios.

\section*{Replication Package}
The replication package is provided in an online repository~\cite{RealityBites}.

\section*{Acknowledgments}
This work is supported by the Co-tester project (No. 314544) and the Co-evolver project (No. 286898/F20), funded by the Research Council of Norway.
Aitor Arrieta is part of the Software and Systems Engineering research group of Mondragon Unibertsitatea (IT1519-22), supported by the Department of Education, Universities and Research of the Basque Country.

\bibliographystyle{unsrt}  
\bibliography{references}

\end{document}